%% file: template.tex
\documentclass{vgtc}                          

\ifpdf
  \pdfoutput=1\relax                   
  \usepackage{graphicx}                
  \DeclareGraphicsExtensions{.pdf,.png,.jpg,.jpeg} 
\else
  \usepackage{graphicx}                
  \DeclareGraphicsExtensions{.eps}     
\fi%

\graphicspath{{figures/}{pictures/}{images/}{./}} 

\usepackage{microtype}                 
\PassOptionsToPackage{warn}{textcomp}  
\usepackage{textcomp}                  
\usepackage{mathptmx}                  
\usepackage{times}                     
\usepackage{cite}                      
\usepackage{tabu}                      
\usepackage{booktabs}                  

\usepackage{xargs}      
\usepackage{soul}       
\usepackage{color}      
\definecolor{lightgreen}{RGB}{101, 184, 101}
\newcommandx{\jane}[2][1=]
    {\setulcolor{lightgreen}{\ul{#1}} \textcolor{lightgreen}
    {[\textbf{Jane:} #2]}}
    
\newcommandx{\rv}[1]
    {{\textcolor{black}
    {#1}}}

\newcommand{\pheading}[1]{\vspace{4px}\noindent\textbf{#1}}
\newcommand{\sysName}{ARShopping}

\onlineid{1092}

\vgtccategory{Application}

\vgtcinsertpkg

\title{ARShopping: In-Store Shopping Decision Support \\ 
Through Augmented Reality and Immersive Visualization}

\author{Bingjie (Jenny) Xu\thanks{e-mail: jenny.xu@u.northwestern.edu}\\
\hspace{-.6cm}
    \scriptsize Northwestern University
\hspace{-.6cm}
\and Shunan Guo\thanks{e-mail: \{sguo, eunyee, jhoffs, rrossi, fdu\}@adobe.com}\\
\hspace{-.6cm}
    \scriptsize Adobe Research
\hspace{-.6cm}
\and Eunyee Koh\footnotemark[2]\\
\hspace{-.6cm}
    \scriptsize Adobe Research
\hspace{-.6cm}
\and Jane Hoffswell\footnotemark[2]\\
\hspace{-.6cm}
    \scriptsize Adobe Research
\hspace{-.6cm}
\and Ryan Rossi\footnotemark[2]\\
\hspace{-.6cm}
    \scriptsize Adobe Research
\hspace{-.6cm}
\and Fan Du\footnotemark[2]\\
    \scriptsize Adobe Research}

\input{sections/00-abstract}

\CCScatlist{
  \CCScatTwelve{Human-centered computing}{Visu\-al\-iza\-tion}{Visu\-al\-iza\-tion systems and tools}{};
  \CCScatTwelve{Human-centered computing}{Human computer interaction (HCI)}{Interaction paradigms}{Mixed / augmented reality}
}

\begin{document}

\maketitle



\input{sections/01-intro}

\input{sections/02-related_work}
\input{sections/03-preliminary}
\input{sections/04-system}
\input{sections/05-visualization}

\input{sections/06-evaluation}
\input{sections/08-conclusion}



\bibliographystyle{abbrv-doi}

\bibliography{reference}
\end{document}

%% file: sections/00-abstract.tex
\abstract{
    Online shopping gives customers boundless options to choose from, backed by extensive product details and customer reviews, all from the comfort of home; yet, no amount of detailed, online information can outweigh the instant gratification and hands-on understanding of a product that is provided by physical stores. However, making purchasing decisions in physical stores can be challenging due to a large number of similar alternatives and limited accessibility of the relevant product information (e.g., features, ratings, and reviews). \rv{In this work, we present ARShopping: a web-based prototype to visually communicate detailed product information from an online setting on portable smart devices (e.g., phones, tablets, glasses)}, within the physical space at the point of purchase. This prototype uses augmented reality (AR) to identify products and display detailed information to help consumers make purchasing decisions that fulfill their needs while decreasing the decision-making time. In particular, we use a data fusion algorithm to improve the precision of the product detection; we then integrate AR visualizations into the scene to facilitate comparisons across multiple products and features. We designed our prototype based on interviews with 14 participants to better understand the utility and ease of use of the prototype.
}

%% file: sections/01-intro.tex
\section{Introduction}Purchasing decisions are often influenced by the product information provided at the point of sale~\cite{tellis1990best}. As online shopping becomes more popular, consumers increasingly rely on the comprehensive and precise product information available online (e.g., detailed product features, historical reviews, price trends, etc.) to make purchasing decisions~\cite{kowatsch2011mobile}. For example, consumers on a diet usually need to review and compare the nutrition information of their groceries to select foods that best fit their nutrition goals. When online shopping, the shopping context and product information can be easily retrieved from the product database and web logs; many prior systems have therefore used this information to help online shoppers make better purchase decisions and provide product recommendations\mbox{~\cite{li2011construction, al2013adaptive, zhong2017poolside}.} The situation of in-store purchases, however, introduces new challenges for retrieving and displaying product information on site.

Recent advancements in augmented reality (AR) provide new opportunities to bridge online product information with physical in-store products, and better communicate information to support comparisons and help shoppers make faster purchasing decisions. Existing work has explored the use of augmented reality in product information retrieval~\cite{ganapathy2011mar, Spreer:2012, gutierrez2018phara}, shopping navigation~\cite{Sayed:2015, Jayananda:2018}, and real-time product recommendation~\cite{ahn2015supporting}, which are all demonstrated to be helpful in improving consumers' in-store shopping experience. However, most existing AR applications focus on detecting one product at a time, while consumers in physical stores are often exposed to many similar product alternatives~\cite{lee2010interaction} and expected to make a reasonable decision between them. In addition, customers may need to consider multiple factors to make purchasing decisions (e.g., nutrition, price, ratings, etc.). Thus, there is a need for future AR systems
to better support decision-making by facilitating comparisons of multiple products across multiple relevant features.

\rv{In this work, we contribute a web-based prototype, ARShopping, for augmenting the in-store shopping experience with online product information (e.g., product features, ratings, etc.) through portable smart devices (e.g., phones, tablets, glasses)}. Inspired by the successful practice of applying data visualization to facilitating multivariate data comparison~\cite{gleicher2011visual, cao2018z}, we present the product information in the form of visual glyphs that can represent multiple features and enable intuitive comparisons across different products. In particular, we developed a product detection algorithm that fuses results from object detection and product marker detection to support better positioning of the glyphs in the space.
We confirmed the needs for in-store decision support and product comparison via interviews with 14 participants, and implemented the prototype based on their feedback. \rv{To evaluate the usefulness and usability of the prototype, we conducted a second interview study with the participants to collect feedback and investigate their experience with using the system prototype in a virtual grocery shopping environment.}

%% file: sections/02-related_work.tex
\section{Related Work}

\pheading{AR + In-Store Shopping.}
Mobile augmented reality contributes to smart retail settings by improving the value for both customers and retailers alike~\cite{Dacko:2017}. Ganapathy~et~al.~\cite{ganapathy2011mar} developed a system to retrieve product information by matching photos taken by the customer to the product image database. Similarly, Spreer~et~al.~\cite{Spreer:2012} utilized mobile AR applications to display product information at the point of sale. Rashid~et~al.~\cite{Rashid:2015} combined AR and RFID to allow users to browse the shelf and search for a given product in an AR interface. While our prototype provides a similar data intervention experience, we allow customers to scan multiple products at the same time and visualize the results to facilitate product comparisons.

Sayed~et~al.~\cite{Sayed:2015} provided situated and abstract information representation with real-time analytical interaction for shopping. Different from our focus on product comparison, this system focuses more on product searching and navigation. Another category of AR shopping systems focuses on recommending products to customers while shopping. For example, Junho~et~al.~\cite{ahn2015supporting} developed a mobile app to recommend and warn shoppers of grocery items in the current isle based on their personal and family health profiles and the nutrition information of the grocery items. The Easy Shopping app developed by Jayananda~et~al.~\cite{Jayananda:2018} demonstrated a similar usage scenario with integrated product navigation. This system directs users to the aisles of the products, and recommends similar items while displaying the relevant health alerts. 
These systems focus on only a single product at a time, whereas our system requires detection of multiple products at the same time that can then have the information extracted, visualized, and compared by the user to facilitate decision-making. 


\pheading{Decision Support for In-Store Shopping.}
Providing information to customers while shopping can influence their purchasing behavior~\cite{kowatsch2010store}. The advancement of Mobile Commerce~\cite{mennecke2002mobile} and Ubiquitous Commerce~\cite{sheng2008experimental} has enabled mobile decision support for in-store shopping. Early implementations introduced location-based agents (e.g., Shopper's Eye~\cite{fano1998shopper}, Impulse~\cite{youll2000impulse}) and sensor network-based agents (e.g., MyGrocer~\cite{kourouthanassis2003developing}) to retrieve product information at the point of purchase. Shoppers may also retrieve data by scanning product barcodes or QR-codes with their phone camera~\cite{kawamura2008wom, van2006mobile}. 

On top of these technologies, decision support systems~(DSS) have been developed to help consumers interact with physical products to retrieve and compare product information~\cite{jeong2009moderating, broeckelmann2008usage}. 
They utilize consumers' requests for specific product attributes to support decision-making.
Similarly, product recommendation agents~(RA) aim to support consumers in choosing the right product among several alternatives and distractions during in-store shopping~\cite{bamis2008mobility, miller2003movielens}. 
Most of the DSSs or RAs are developed with a mobile user interface, separate from the shopping environment. Recent advancements in eye-tracking~\cite{pfeiffer2015towards} and AR~\cite{gutierrez2018phara} for shopping have enabled new opportunities to provide more context-aware decision support, which could have a large impact on consumers' decision strategy~\cite{bettman1998constructive}.
While mobile shopping agents were proved to be generally useful in prior work~\cite{miller2003movielens}, there is a research trend highlighting the importance of understanding how mobile services may impact consumers' shopping experience and decision-making~\cite{kalnikaite2013decision}. In this paper, we incorporate comparative data visualization into AR for shopping decision support and build an understanding of consumers' potential acceptance towards an AR-based shopping decision support system.

%% file: sections/03-preliminary.tex
\section{Formative Interviews}
\label{sec:formative}
To understand the general needs of in-store customers and how augmented reality can help improve the in-store shopping experience and decision-making, we conducted preliminary interviews with 14 participants (P1--14, 6 males, 8 females, aged 23 to 52). \rv{Participants were recruited via advertisement on social media platform with backgrounds ranging from graduate student, teacher, software engineer, research scientist and speech-language pathologist, under the condition that they had recent experience shopping both online and in-store.}
All interviews were conducted remotely and recorded; each interview took about 30 minutes. 
The interviews were semi-structured covering questions including their shopping frequency and habits, purchasing categories, and key product information for making purchasing decisions. Based on the feedback from participants, we identify the following key insights and further infer the needs for in-store shopping assistance.

\pheading{Shopping Habits and Categories.} We found the product categories that people chose to purchase in-store are relatively fixed and they have a regular frequency of visiting physical stores. For example,  nearly all interviewees (13/14) described a regular in-store cadence for grocery shopping (1--4 times per week). In contrast, people have more diverse shopping categories and flexible frequency online. 

When deciding between in-store or online shopping, the most commonly mentioned factors are the product categories and urgency of the need. In particular, participants usually choose to buy fresh goods (13/14) and household essentials (7/14 participants) in-store. Three interviewees also mentioned that they would visit physical stores for products they found online but would like to try out in person, such as clothes, footwear, glasses, and instruments. Participants would also choose to buy products in urgent need, such as medicine or other necessities through physical stores. One participant noted that differences in price can contribute to the decision of whether to shop in-store or online: \emph{``Sometimes the brick-and-mortar store offers better price to compete with online store''}~(P12). These results suggest  that even with the rapid evolution of e-commerce, people still have regular demand for visiting brick-and-mortar stores, especially for products that are fresh and in urgent need.

\pheading{Pain Points of In-Store Shopping.} When asked about the difficulties participants experienced during in-store shopping, the most common ones include travel time and cost, limited product variety, and the difficulty of making purchasing decisions. In particular, three participants 
mentioned that they may need to \emph{``visit multiple stores due to the lack of supply''}~(P8, 13, 14) and two mentioned that \emph{``it takes multiple days decide among products in different stores''}~(P10, 11). 
P1 described difficulty deciding between different products, often resulting in a less than satisfactory outcome: \emph{``I sometimes need to pickup stuffs on my way back home that my family need immediately with a short notice. It's difficult to choose the best one without researching beforehand so I often have to pick up a random one.''} 
P6 described a similar experience: \emph{``It's hard to determine which brand to buy when purchasing household goods that I do not often use.''} 
P13 also recalled his experience of searching for product information online while purchasing it in-store: \emph{``I had to search the product online with my mobile phone to compare the reviews of different brand.''}
While the cost of travel are the natural defects of in-store shopping, the hardship of making decisions when shopping in-store raises a need for providing decision assistance support in brick-and-mortar stores.

\pheading{Shopping Decision-Making.} We asked participants to recall a recent shopping experience and describe how they decided on the product(s) they purchased. In these conversations, we noticed several key pieces of information that they often referred to when making decisions, including the composition, efficiency, historical reviews, prices, and discounts across platforms. For example, P1 recalled his recent experience purchasing medications
and noted that he referred to the chemical composition and historical reviews for different brands. 
Similarly, P5 recalled her habit of reading nutrition labels when purchasing snacks. 
While this information is easily accessible online, oftentimes it is not available or is more difficult to browse when customers are purchasing in-store. 
In fact, seven participants brought up searching online for this information before visiting a physical store, and five participants mentioned a pressing need to be able to conduct this research  while being in-store in order to decide which one to buy. 
For example, P9 said that \emph{``There were several times when I could not decide which one to buy in-store and also felt inconvenient to research on my phone, so I took a picture of all the available brand and returned home to look them up.''}
These results reinforce the need for providing decision assistance in-store that can better enable customers to access and compare relevant online information for the products immediately available.

%% file: sections/04-system.tex
\section{\sysName\ Prototype}
\begin{figure}[t!]
\includegraphics[width=\linewidth]{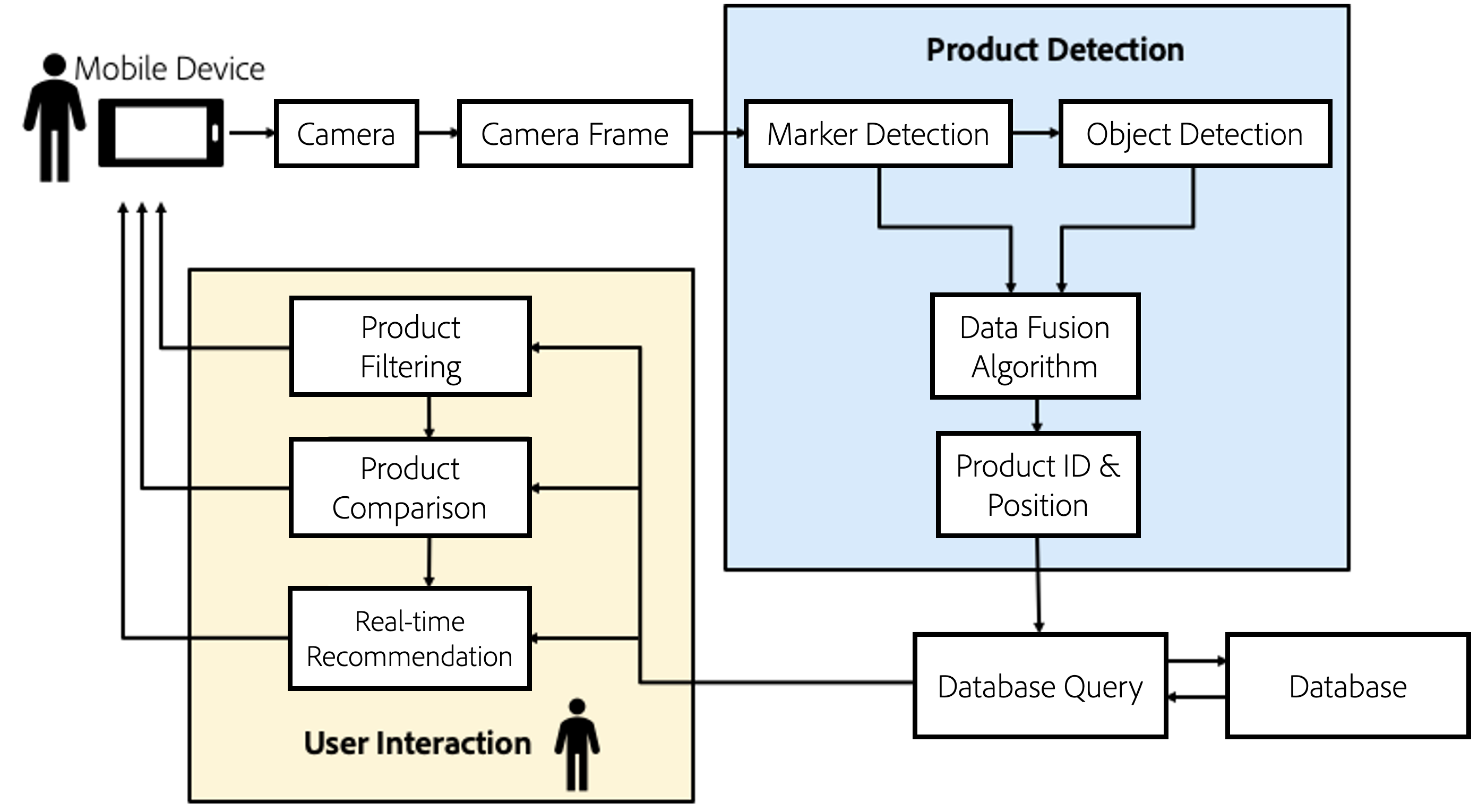}
\vspace{-0.5cm}
\caption{The architecture of the ARShopping prototype.}
\vspace{-0.4cm}
\label{fig:system}
\centering
\end{figure}

Our system, \sysName, includes two modules (as shown in Fig.~\ref{fig:system}): a product detection module and a user interaction module with the mobile interface. 
The product detection module~(Sec.~\ref{sec:detection}) retrieves the product ID and position by leveraging both the marker that is attached to the product price tag and the image of the products in the scene captured from the mobile camera. 
The product detection module also identifies the 3D position, rotation, and actual size of each product for the placement of our comparative visualization. 
The user interaction module~(Sec.~\ref{sec:vis}) includes the main functionality for the user interface, including the comparative visualizations. 

\subsection{Product Detection}
\label{sec:detection}
As our prototype needs to include multiple products in the scene and detect them at the same time, there may be mutual interference between different products when they are jammed on the shelf. To enhance the stability and applicability of the detection, we incorporate an additional marker with each product and combine the results of marker detection and product image detection. 

\pheading{Implementation.} 
\rv{For marker detection, we utilized the built-in marker tracking function from the \textit{AR.js} library. We print a marker with a unique ID encoded for each product and assume the marker will be placed close to the corresponding physical product (e.g., on the price tag) and also on the same surface of the product shelf. In this way, we can infer the product ID based on the encoded information and the 3D position, rotation, and actual size of the physical product based on the detected size, transformation, and rotation of the marker. The product object detection is implemented based on an object detection model trained by online images of each product and transfer learning from the TensorFlow object detection model \cite{tan2020efficientdet}. For each object detected, the model outputs the matched product ID, product's center position, and the size of the bounding box. To reduce the overlap of the immersive data visualizations for multiple products~(Sec.~\ref{sec:vis}), it is necessary to properly layout the visualization as users interact with the system. Therefore, we further determine the layout of the visualization by finding the closest product near the marker and use the area of the product to overlay the visualization. }The objects’ locations are transferred from screen coordinates to world coordinates to calculate their distance to the marker, assuming that the product is at the same distance as the marker from the camera.

\pheading{Online Database Query.}
With the detected product IDs, we can retrieve the online information of
products in the camera scene from the online database. 
For this initial prototype, we crawl grocery information from Target.com to establish our database, which includes the price, ratings, in-store deals/coupons, and specifications (e.g.,~nutrition, size, weight, etc.) of each product. After detecting products through the camera, the system sends the corresponding product IDs in the scene to the online database to query the product information. At the same time, the system will also search the database to see if the products are on sale or has any in-store coupons that can be applied, and push the coupon to users in real-time.

\pheading{Data Fusion Algorithm.}
For each product, we utilize a data fusion algorithm to determine its existence and position.
Note that products' position refers to a broad definition of position, including the product's center position, size, transformation, and rotation. There are four different cases for each product: (1)~When both methods have detected this product, its center position and size are from the object detection results and its transformation and rotation are from the marker detection, assuming the product's front surface is parallel to the marker. (2)~When the product has only been detected by the marker detection, its position is calculated only from marker's center position, size, transformation, and rotation with a fixed relative position between the marker and the product (e.g., the product is on the top of the marker) and a fixed scale between the marker's size and the product's size. (3)~When the product was only detected by object detection, its position only contains the center position and size from the object detection result assuming it has no transformation and rotation relative to the camera. (4)~When both methods have not detected this product, it is considered to not exist in the scene.




%% file: sections/05-visualization.tex
\subsection{User Interface}
\label{sec:vis}

\begin{figure}[t!]
\includegraphics[width=\linewidth]{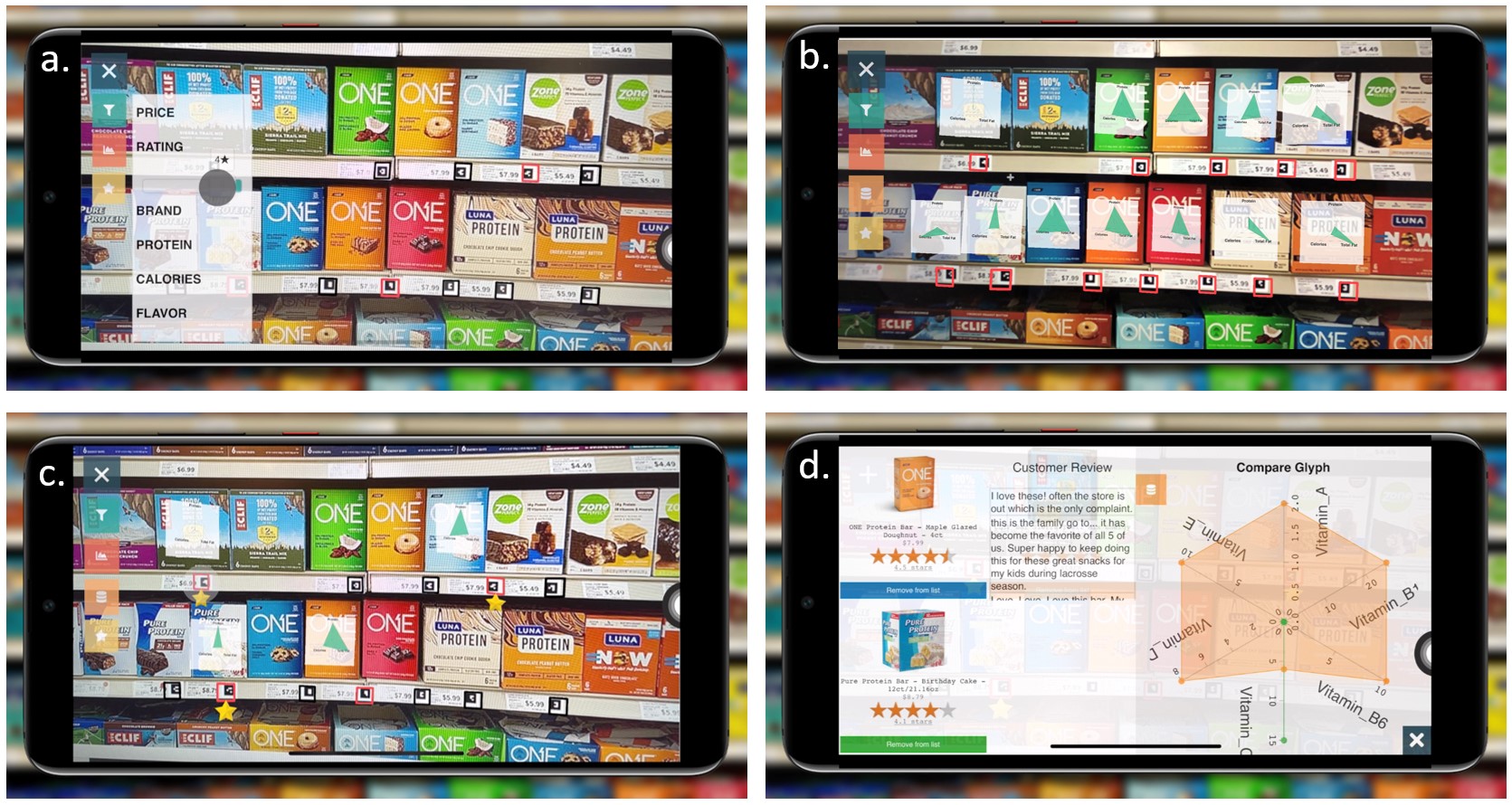}
\vspace{-15px}
\caption{The ARShopping Interface. 
(a) Filtering. 
(b) Multivariate data glyphs. 
(c) Adding products to favorites.
(d) Comparison view.}
\label{fig:visulization}
\centering
\end{figure}


\noindent\textbf{Visualizing product information.}
ARShopping utilizes radar charts to represent multivariate product data. After preliminary screening and filtering (Fig. \ref{fig:visulization}a), users can analyze qualified products' information by visualizing glyphs overlaid on the physical products~(Fig.~\ref{fig:visulization}b). The layout of the glyph is calculated based on the resulting product position determined by our product detection method. 
For \rv{marker} detection only, the glyphs may experience some overlap and visually clutter the space due to the placement of the marker (Fig.~\ref{fig:glyph}a). By leveraging the object detection, we can create a more natural integration of the physical product and the virtual marker. In particular, we leverage the orientation of the marker as read from the camera to tell the orientation of the customer. The canvas of the glyphs will be distorted to better reflect the customer's point of focus (Fig.~\ref{fig:glyph}b). 

Regardless of which method is employed, our glyph placement algorithm creates a direct mapping between products and the critical online information necessary to make an informed purchasing decision; this way, customers can find information appropriately situated in the space and easily select the desired product once a purchasing decision is made. 
Glyphs can be triggered on and off by clicking the glyph button on the menu. When glyphs are shown, users can decide what information to show by clicking the database button and selecting information of interest on the dropdown list next to the button. Each axis of the radar chart has the same scale that considers all products of the same type in the store, which assists users in comparing products' information by observing the relative amount of products' specific data intuitively.

\pheading{Comparison view.}
The main view presents the glyphs in juxtaposition, which gives an overview of the features across all products. However, it may hinder customers from finding the optimal one when mutliple products have similar features. To further help customers make final purchase decision, ARShopping also enables users to make more detailed comparisons via the \textit{comparison view}~(Fig.~\ref{fig:visulization}d), in which the glyphs are placed in superposition for more thorough feature comparison. Users can add products they want to compare to the favorite list (Fig.~\ref{fig:visulization}c), and the \textit{comparison view} will show the overlaid radar charts for feature comparison with subjective features such as ratings and customer reviews.

%% file: sections/06-evaluation.tex
\begin{figure}[t!]
\includegraphics[width=\linewidth]{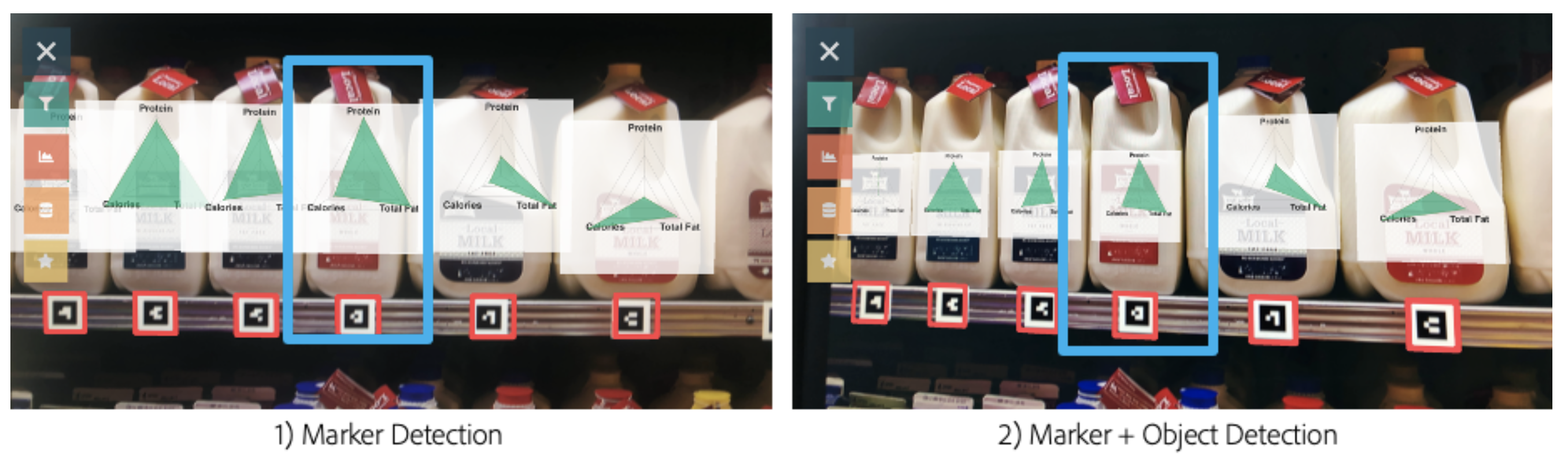}
\vspace{-20px}
\caption{A comparison of product detection and glyph placement results with (a) marker detection only and (b) marker+object detection.}
\vspace{-0.3cm}
\label{fig:glyph}
\centering
\end{figure}

\section{Evaluation of \sysName}
To assess the usability and usefulness of our \sysName\ prototype for assisting with decision making during in-store shopping, we conducted a second round interview through a case study \rv{in a virtual grocery shopping setting} with the 14 participants we recruited from the formative interviews (Sec.~\ref{sec:formative}). 

\subsection{Study Procedure}
In each interview, we first demonstrated the prototype's usage and functionalities through an introduction video with a usage scenario of purchasing protein bars in the grocery store. The introduction video is included in the Supplemental Material. After that, we sent the participant a picture of milk bottles on shelves in the grocery store, which was preprocessed with a marker attached next to the price tag of each product. We asked the participant to display the picture in full screen on a laptop. The prototype was deployed on a remote web server, and the participants could access it through a web link with the browser on their phone. We asked the participants to scan the picture with the prototype as if they were purchasing milk in the grocery stores and freely explore the prototype's functionalities. Participants were encouraged to think out loud, give feedback, and raise questions about the prototype. After the prototype try-out, we further interviewed the participants to collect their subjective feedback in terms of the usefulness, ease of use, and directions for enhancing the current prototype. In the end, we asked participants to answer a post-study questionnaire with eight questions (shown in Fig.~\ref{fig:survey}) on a 7-point Likert scale designed following the Technology Acceptance Model (TAM)
to measure participants' perceived usefulness (Q1--Q4), ease of use (Q5--Q7), and intention of use (Q8). 

\subsection{Acceptance}
\label{sec:acceptance}
\begin{figure}[t!]
\includegraphics[width=\columnwidth]{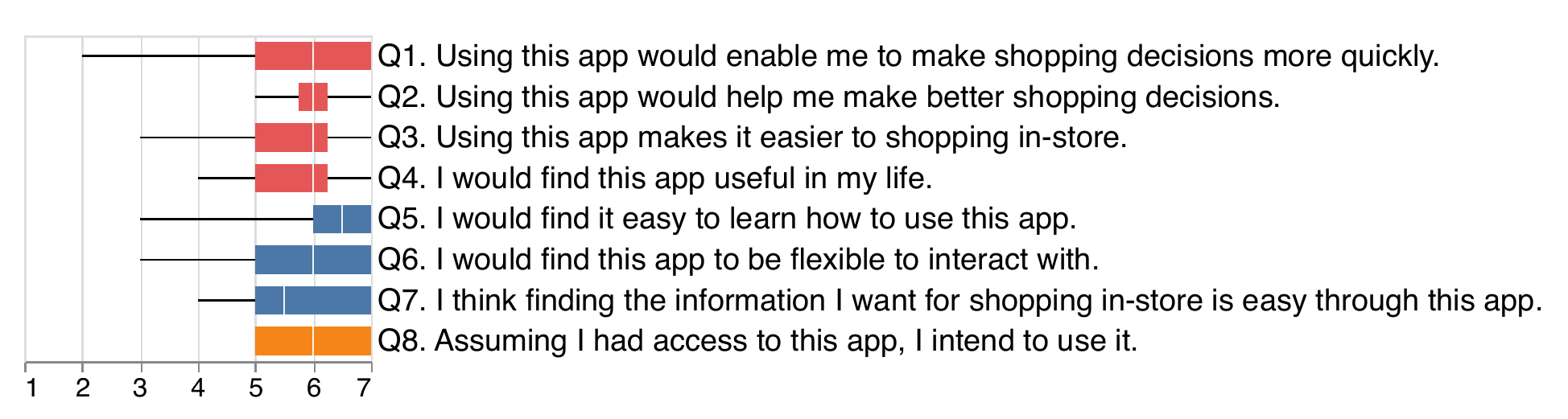}
\vspace{-20px}
\caption{Post-study survey results with Technology Acceptance Model (TAM) instruments. Responses were made on a 7-point Likert scale (1=``strongly disagree'', 7=``strongly agree''). Q1--Q4 explore the perceived usefulness (in red), Q5--Q7 explore the perceived ease of use (in blue), and Q8 asks about intention of use (in orange).}
\vspace{-0.3cm}
\label{fig:survey}
\centering
\end{figure}

As shown in Fig.~\ref{fig:survey}, participants generally agreed the prototype was useful and easy to use, and also showed their willingness to use the prototype in their daily life. Specifically, the average perceived usefulness, ease of use, and intention of use of all responses were \textit{5.78}, \textit{5.45}, and \textit{6.07}, respectively. For the usefulness and ease of use, we also tested the reliability of the responses with Cronbach's alpha scoring \textit{0.682} and \textit{0.881}, respectively, which indicates that the observed acceptance towards the prototype is reliable ($>0.6$).

\subsection{
Feedback and Implications}
\label{sec:feedback}


\pheading{Perceived usefulness.} As reported in Sec.~\ref{sec:acceptance}, participants generally found the prototype to be useful. In particular, seven participants commented that this prototype can \emph{``save their time of comparing''} (P1, 5, 7, 8, 12, 13, 14). Three participants felt the prototype can help them \emph{``get to know the product better''} and it can be particularly useful when they are  \emph{``making purchasing decisions on products they are not familiar with''} while \emph{``increasing the confidence of purchasing''} (P2, 10, 14). 
For example, one participant noted that \emph{``This prototype provides a data-driven perspective which is more trustworthy than other people's opinion''} (P10).
In terms of which feature participants felt was most useful, twelve participants voted for the comparative visualization and six voted for the overall filtering functionality. 
As P9 said: \emph{``I really like how filtering and comparative visualization work together. As the filtering roughly set the range of my purchasing options, the comparative visualization can provide decision support in much finer detail.''} 

While many participants were positive about the utility of the system for decision-making, several participants raised concerns about
the usefulness of integrating data-driven comparisons. For example, P4 noted that \emph{``Showing data is probably not friendly enough for novice users. I would need more text description of the product to make decisions.''} Other participants remained more neutral about the usefulness of the prototype, as they think it may depend on the \emph{``value of the product''}~(P3), \emph{``precision of the scanning''}~(P7, 8, 9), and the \emph{``familiarity of the product''}~(P2, 10). 

\pheading{Ease of use.} Participants did not experience much difficulty when trying out the prototype, which could be due to the simplicity of the prototype in the current stage. While most participants appreciated that the prototype was a web-based application that could eliminate the need to download and install a separate application, two participants\rv{~(P8, 14)} raised their concern about the internet connection in physical stores and wondered if the prototype can be used offline. One participant (P7) noted that the prototype might consume a lot of phone battery as the camera needs to be constantly turned on. P8 also brought up that the comparative visualization may be difficult to investigate when the number of parameters is large. 

\pheading{Desired features.} We asked participants about the functionalities they wished to have besides the features of the prototype. Many of them brought up product recommendation, either by a predefined user need or the purchases of other relevant customers. Participants also expressed a desire to have price comparisons both online and offline across purchase channels, and to show the historical trends of the price. P4 pointed out that the application could use more text descriptions to make the data-driven analysis and visualization more friendly to novice users. P14 also suggested to incorporate more text descriptions to explain the meaning of the product specifications.

\section{Limitations and Future Work}
\rv{While our prototype is implemented as a web application and can be used on any smart devices with a camera, we only tested its usage on smartphones, leaving the use on tablets and smart glasses for future investigation. }In addition, since this study was conducted during the pandemic, we decided to employ the virtual setting, which may have covered up some potential usability issues that could arise during real-world usage. For example, there could be some interference in a real-world environment that would affect the stability of the product identification, such as natural light and shadows, and the size of the markers and products in the scene. In fact, our proposed product detection solution was partially motivated by these issues~(i.e.,~combining marker detection and object detection results to improve detection precision). However, the real-world experience and performance still needs to be evaluated in future work.






%% file: sections/08-conclusion.tex
\section{Conclusion}
ARShopping provides new possibilities for a better in-store shopping experience augmented with data visualizations and online product details. 
Compared with existing in-store shopping AR applications, ARShopping aims to help consumers make better purchasing decisions across multiple similar alternatives. To enhance the stability and precision of the product detection, we propose to incorporate marker detection and object detection for product information retrieval. We also integrate data visualization to help with displaying and comparing comprehensive product specifications. Our interview studies with consumers point out several future research directions, including comparing price throughout time and across different purchase channels, improving the design of visualization to make it more understandable to novice users, and incorporating more narratives to make the comparison results more intuitive.
